\documentclass[sigconf]{acmart}
\AtBeginDocument{%
  }

\copyrightyear{2026}
\acmYear{2026}
\setcopyright{rightsretained}
\acmConference[CHI EA '26]{Extended Abstracts of the 2026 CHI Conference on Human Factors in Computing Systems}{April 13--17, 2026}{Barcelona, Spain}
\acmBooktitle{Extended Abstracts of the 2026 CHI Conference on Human Factors in Computing Systems (CHI EA '26), April 13--17, 2026, Barcelona, Spain}\acmDOI{10.1145/3772363.3778744}
\acmISBN{979-8-4007-2281-3/2026/04}

\usepackage{booktabs}
\usepackage{multirow}
\usepackage{graphicx}
\usepackage{array}

\begin{document}


\title{From Human-Human Collaboration to Human-Agent Collaboration: A Vision, Design Philosophy, and an Empirical Framework for Achieving Successful Partnerships Between Humans and LLM Agents}


\author{Bingsheng Yao}
\email{b.yao@northeastern.edu}
\affiliation{%
  \institution{Northeastern University}
  \city{Boston}
  \state{Massachusetts}
  \country{USA}
}

\author{Chaoran Chen}
\email{cchen25@nd.edu}
\affiliation{
  \institution{University of Notre Dame}
  \city{Notre Dame}
  \state{Indiana}
  \country{USA}
  }

\author{April Yi Wang}
\email{april.wang@inf.ethz.ch}
\affiliation{
  \institution{ETH Zurich}
  \city{Zurich}
  \country{Switzerland}
  }


\author{Sherry Tongshuang Wu}
\email{sherryw@cs.cmu.edu}
\affiliation{
 \institution{Carnegie Mellon University}
  \city{Pittsburgh}
  \state{Pennsylvania}
  \country{USA}
}

\author{Toby Jia-jun Li}
\email{toby.j.li@nd.edu}
\affiliation{
  \institution{University of Notre Dame}
  \city{Notre Dame}
  \state{Indiana}
  \country{USA}
  }

\author{Dakuo Wang}
\email{d.wang@northeastern.edu}
\affiliation{%
  \institution{Northeastern University}
  \city{Boston}
  \state{Massachusetts}
  \country{USA}
}

\renewcommand{\shortauthors}{Yao et al.}
\renewcommand{\shorttitle}{From Human-Human Collaboration to Human-Agent Collaboration}
\begin{abstract}

The emergence of Large Language Model (LLM) agents enables us to build agent-based intelligent systems that move beyond the role of a ``tool'' to become genuine collaborators with humans, thereby realizing a novel human-agent collaboration paradigm. 
Our vision is that LLM agents should resemble remote human collaborators, which allows HCI researchers to ground the future exploration in decades of research on trust, awareness, and common ground in remote human collaboration, while also revealing the unique opportunities and challenges that emerge when one or more partners are AI agents.
This workshop\footnote{Workshop website: \href{https://chi26workshop-human-agent-collaboration.hailab.io/}{https://chi26workshop-human-agent-collaboration.hailab.io/}} establishes a foundational research agenda for the new era by posing the question: How can the rich understanding of remote human collaboration inspire and inform the design and study of human-agent collaboration? 
We will bring together an interdisciplinary group from HCI, CSCW, and AI to explore this critical transition. 
The 180-minute workshop will be highly interactive, featuring a keynote speaker, a series of invited lightning talks, and an exploratory group design session where participants will storyboard novel paradigms of human-agent partnership.
Our goal is to enlighten the research community by cultivating a shared vocabulary and producing a research agenda that charts the future of collaborative agents.

\end{abstract}

\ccsdesc[500]{Human-centered computing~Human computer interaction (HCI)}
\ccsdesc[500]{Human-centered computing~Interaction paradigms}
\ccsdesc[500]{Human-centered computing~Collaborative interaction}

\keywords{human-agent collaboration, human-AI collaboration, remote collaboration, large language model, LLM agent}
\begin{teaserfigure}
  \includegraphics[width=\textwidth]{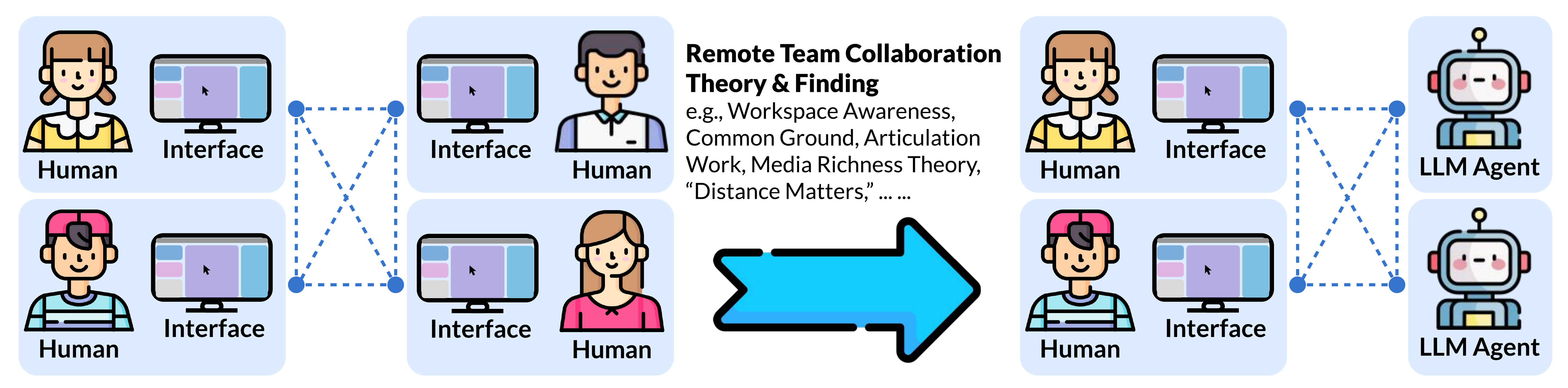}
  \caption{Our vision, design philosophy, and empirical framework: a novel research agenda from human-human collaboration to human-agent collaboration. }
  \label{fig:teaser}
\end{teaserfigure}


\maketitle

\section{Motivation and Goals}


Computing has long been shaped by paradigm shifts that redefine the relationship between humans and intelligent systems~\cite{grudin1988cscw,grudin2018tool}.
A foundational challenge in Human-Computer Interaction (HCI) and Computer-Supported Cooperative Work (CSCW) is moving beyond the ``tool'' paradigm, where systems act as passive instruments awaiting human instructions, to interactions that resemble genuine ``teammate''~\cite{grudin1988cscw, horvitz1999principles,grudin2018tool, amershi2019guidelines}. 
Systems that function as tools often lack the critical characteristics of effective human partnerships, such as mutual awareness~\cite{dourish1992portholes, harrison1996re, gutwin2002descriptive} (a shared understanding of who is doing what), adaptivity~\cite{horvitz1999principles} (the ability to dynamically adjust to a partner's changing goals), and shared accountability~\cite{kroll2015accountable} (a joint responsibility for outcomes). 
Without these qualities, interactions between humans and systems can be brittle and error-prone, making trust calibration difficult, and the potential for synergistic collaboration remains largely unrealized~\cite{lee2004trust,amershi2019guidelines}.

Recent advances in Large Language Models (LLMs) mark a new inflection point. 
\textbf{LLM agents} can sustain open-ended natural language dialogue, exhibit reasoning and planning, and even evoke social or moral responses from users through role-playing~\cite{park2023generative, leong2024dittos, wu2024autogen, ye2025my, yao2025dprf, chen2025multi, lu2025llm}. 
At the same time, users often project emotions, empathy, and even moral expectations onto these systems~\cite{reeves1996media,epley2007seeing,10.1145/3613904.3642363}, suggesting that interactions with LLM agents increasingly resemble partnerships rather than tool use.
These dynamics suggest a shift toward \textbf{mixed-initiative collaboration}~\cite{horvitz1999principles}, where agents proactively coordinate, maintain common ground, and negotiate labor~\cite{10.1145/3746059.3747798}. 
Yet, everyday failures already reveal the risks: code assistants that conflict with ongoing edits, writing tools that summarize without provenance, and hallucinations that silently undermine reliability~\cite{amershi2019guidelines,ji2023towards}. 
Without careful design, such failures could entrench tool-like brittleness and introduce new harms~\cite{chen2025obvious, tang2025dark, gebreegziabher2026behavioral}. 
Our workshop is thus motivated by the urgent need to establish a principled, research-driven foundation for navigating this dual landscape of promise and risk.

We propose a \textbf{novel human-AI interaction design philosophy}: to design and study \textbf{LLM agents as remote human collaborators}~\cite{yao2025through}. 
This analogy grounds human-agent collaboration in decades of CSCW research on distributed teamwork~\cite{olson2000distance,cramton2001mutual,clark1991grounding}.
In remote collaboration, humans communicate and coordinate through lean channels, lack shared physical context, and must rely on explicit signals to maintain common ground~\cite{daft1986organizational,clark1991grounding}, which is analogous to how humans interact with LLM agents.
Adopting this vision raises various design questions:
How can humans establish \textbf{common ground} with agents that lack lived experience~\citet{clark1991grounding}? What representations support \textbf{workspace awareness} of an agent’s focus and intent~\cite{gutwin2002descriptive}? When should agents take on \textbf{articulation work}, and how should they expose it for oversight~\cite{schmidt1992taking}? What forms of \textbf{media richness} enable effective negotiation despite lean communication~\cite{daft1986organizational}?

This vision structures our workshop around a dual inquiry: \textbf{foundations} and \textbf{frontiers}.
Foundations involve re-examining CSCW theories of trust, awareness, and coordination to provide theoretical rigor for human–agent interaction~\cite{olson2000distance,gutwin2002descriptive,lee2004trust}. 
Frontiers involve identifying where the existing principles no longer stand and new paradigms arise, such as asynchronous collaboration (agents working in a user's absence) and multi-human, multi-agent teamwork with complex organizational structure~\cite{wu2024autogen,park2023generative}.

We also acknowledge the limitations and boundaries of our ``remote collaborator'' vision and design philosophy. 
Agents lack embodiment, lived experience, and social accountability, and their errors manifest in unfamiliar forms such as fabricated citations, privacy leaks, value misalignments, or cascading failures~\cite{carlini2021extracting,ji2023towards, 10.1145/3708359.3712156, chen2025obvious,tang2025dark,gebreegziabher2026behavioral}.
Rather than shortcomings, we view them as design challenges: how might we make failures legible, safeguard privacy, and build accountability when collaborators are not human~\cite{kroll2015accountable,chen-etal-2025-towards-design,chen2025toward,10.1145/3733816.3760757}?
We also draw on the ``beyond being there'' argument to focus on benefits that are hard in human-only teams, such as persistent memory with provenance, machine-readable task contracts, and programmable social protocols~\cite{hollan1992beyond}.
The ``remote collaborator'' lens is therefore not a perfect mapping but a generative heuristic: it highlights where CSCW theories offer guidance while also signaling where entirely new frameworks are needed.

This workshop provides the first dedicated forum for the HCI community to collectively engage with this dual inquiry. 
We aim to create a shared vocabulary and a research agenda that is both grounded in foundational knowledge and visionary about future possibilities~\cite{amershi2019guidelines}.
Our objectives are:
\begin{enumerate}
    \item Establish the ``agent as a remote collaborator'' analogy as a bounded but powerful heuristic for research.
    \item Assess the applicability of the core CSCW theories to emerging human-agent contexts.
    \item Identify and ideate novel interaction paradigms that extend beyond traditional human-human models.
    \item Integrate considerations of risks (security, privacy, trust miscalibration) into the design space of collaboration.
    \item Develop a community-driven research agenda that is suitable for the future of human-agent collaboration studies.
\end{enumerate}

\section{Organizers}

The organizers of this workshop are interdisciplinary researchers in HCI, CSCW, and NLP with work on agent-mediated collaboration, mixed-initiative systems, privacy, and accountability.
Our record includes leading CHI and CSCW publications, prior workshops organizations, the release of open resources for agent research, and empirical studies in software engineering, writing support, healthcare, and education. The workshop agenda grows directly from this experience and from recent HCI literature on human–agent collaboration, which together motivate the scope, methods, and expected outcomes of the event.

\textbf{Bingsheng Yao} is an Associate Research Scientist at Northeastern University's Khoury College of Computer Sciences. 
His research aims to facilitate effective human-AI collaboration in domain-specific scenarios through the design and development of human-centered NLP systems, and to investigate emerging human-agent collaboration paradigms in which humans collaborate with LLM agents in mutual partnerships. 
He has recently co-organized the ACM CHI 2024 Special Interest Group on LLM privacy and a workshop at CSCW 2024 on LLM-based synthetic personae.

\textbf{Chaoran Chen} is a fourth-year Ph.D. Candidate in the Department of Computer Science and Engineering at the University of Notre Dame. His research uses human-centered design to investigate both the risks introduced by LLM agents and their potential as tools for improving user privacy and security. His work emphasizes the role of novel interaction and interface designs in empowering end users to navigate the evolving privacy and security landscape shaped by generative AI. He has served on the program committee of the 1st Workshop on Human-Centered AI Privacy and Security at ACM CCS 2025 and actively contributes to the HCI and usable security communities, with first-author publications at top-tier HCI and Security conferences.

\textbf{April Wang} is an Assistant Professor in the Department of Computer Science at ETH Zurich, where she leads the Programming, Education, and Computer–Human Interaction (PEACH) Lab.
Her research in HCI and educational technology designs intelligent, literate programming and notebook-based environments that improve communication, collaboration, and learning in programming, including human–AI collaboration.
She has served on the organizing committee and program committee of various venues, including VL/HCC, CHI, UIST, and L@S.


\textbf{Sherry Tongshuang Wu} is an Assistant Professor at the Human-Computer Interaction Institute, Carnegie Mellon University. Her primary research investigates how humans (AI experts, domain experts, and lay users) interact with (debug, audit, and collaborate) AI systems. Sherry has organized three workshops at NLP and HCI conferences: Shared Stories and Lessons Learned Workshop at EMNLP 2022 and Trust and Reliance in AI-Human Teams at CHI 2023-2024. She has also given three tutorials related to Human-AI Interaction at EMNLP 2023, NAACL 2024, and ACL 2025.

\textbf{Toby Jia-Jun Li} is an Assistant Professor in the Department of Computer Science and Engineering at the University of Notre Dame, where he leads the SaNDwich Lab. He also serves as the Director of the Human-Centered Responsible AI Lab at the Lucy Family Institute for Data \& Society. He and his group use human-centered methods to design, build, and study human-AI collaborative systems. The group's projects have been generously supported by NSF, Google, AnalytiXIN, and other agencies. He has successfully organized workshops at multiple conferences, including CHI, CSCW, and IUI. He has also served on various organizing committees, including as the General Chair for ACM IUI 2025.

\textbf{Dakuo Wang} is an Associate Professor at Northeastern University, jointly appointed at Khoury College of Computer Sciences and the College of Arts, Media, and Design. His research lies at the intersection of human-computer interaction (HCI) and artificial intelligence (AI), with a focus on the exploration, development, and evaluation of human-centered AI (HCAI) systems. In addition to his research interest in human-AI collaboration and collaborative editing, he has been active in multiple steering committees for the HCI community, such as CHI 2019 Conference Co-Chair in Social Media, ACM CHI 2022 Conference Co-Chair in Special Interest Group, and CHI 2023 Conference Co-Chair in Workshop.

\section{Workshop Structure and Activities}

\begin{table*}[]
\caption{Tentative schedule for the 180-minute workshop. The workshop combines knowledge-sharing (keynote, panel discussion, lightning talks, spotlight papers, poster session) with interactive group activities (Session 2).}
\vspace{-0.5em}
\label{tab:timeline}
\resizebox{\textwidth}{!}{%

\begin{tabular}{@{}%
  >{\raggedright\arraybackslash}p{0.06\textwidth}%
  >{\raggedright\arraybackslash}p{0.16\textwidth}%
  >{\raggedright\arraybackslash}p{0.78\textwidth}@{}}

\toprule
       & \textbf{Time Slot} & \textbf{Activities}                 \\ \midrule
\multirow{13}{*}{Session 1} & 1:30 pm–1:35 pm      & \textbf{Introduction}                        \\
& 1:35 pm–2:05 pm     & \textbf{Keynote with Q\&A} (Diyi Yang, Stanford University, \textit{tentative}) \\
& \multirow[t]{2}{*}{2:05 pm–2:25 pm}     & 
\begin{minipage}[t]{\linewidth}
\textbf{Panel Discussion} \\
Panelist (\textit{tentative}): Dakuo Wang, Toby Jiajun Li, April Wang  
\end{minipage} \\

& \multirow[t]{7}{*}{2:25 pm–3:00 pm} &
\begin{minipage}[t]{\linewidth}
\textbf{Featured Lightning Talks:}

\textit{1. Through the Lens of Human-Human Collaboration: A Configurable Research Platform for Exploring Human-Agent Collaboration} (Bingsheng Yao, Northeastern University)\\
\textit{2. Dark Patterns Meet GUI Agents: LLM Agent Susceptibility to Manipulative Interfaces and the Role of Human Oversight} (Chaoran Chen, University of Notre Dame)\\
\textit{3. ...} \\
\textit{4-6. ... (Authors of three to four selected workshop submissions will present their work.) } 
\end{minipage} \\


\midrule
& 3:00 pm-3:30 pm & \textbf{Coffee Break (Concurrent with poster session)} \\
\midrule
\multirow{5}{*}{Session 2} & 3:30 pm-4:30 pm  & 
\begin{minipage}[t]{\linewidth}
\textbf{Exploratory Design Sessions:} \\
Consists of two 30-minute sessions. During each session, each group receives a scenario probe centered on a collaborative challenge re-imagined for human-agent collaboration. Storyboard a critical moment and focus on the interaction designs and research considerations. 
\end{minipage} \\
                           & 4:30 pm-5:00 pm & \textbf{Group Presentations, Synthesis, and Closing} \\ 
\bottomrule
\end{tabular}%
}
\vspace{-1em}
\end{table*}

Our workshop will be a \textbf{Long (180-minute)}, fully in-person event with a target audience population of \textbf{40-50 participants}. It is designed to provide theoretical framing, showcase diverse perspectives, and facilitate interactive design activities.

\subsection{Pre-Workshop Plan}

Participants will submit a short paper (2-4 pages, ACM single-column format).
We welcome ongoing research, survey papers, and position papers that stimulate discussion on human–agent collaboration.
Submissions will be reviewed by two program committee members and one organizer. 
Five papers will be selected as spotlights, with authors giving 3-minute spotlight talks during the workshop.
The program committee will include experts across HCI, NLP, and social sciences, who will also help promote the workshop.
We are committed to diversity, equity, and inclusion, and particularly encourage participation from underrepresented groups and those studying their needs.


\subsection{Communication and Advertising Channels}

We will maintain multiple communication channels, including a \textbf{website}, a \textbf{mailing list}, and a dedicated \textbf{Slack workspace}, allowing participants to be aware of the latest updates before, during, and after the workshop.
The website will provide all critical information for our workshop, including the Call for Participation, submission instructions, program schedule, and contact information for the program committee. 
The website will also be updated regularly to advertise and disseminate workshop contributions, including accepted papers, featured research, and a post-workshop plan for a white paper (details in Section~\ref{sec:post-workshop}). 
In addition, we will advertise our workshop through the \textbf{mailing list of research institutions and related conferences} in interdisciplinary research (e.g., IUI, CHI, CSCW, ACL), as well as \textbf{direct communication} with collaborators and colleagues through social media and in-person meetings.

\subsection{Workshop Schedules}

Table~\ref{tab:timeline} presents the schedule of our proposed workshop, which spans two sessions with a total of 180 minutes.
We will adjust the acceptance rate and structure accordingly, depending on the submission amount. For instance, if many more applicants than expected show interest in participating after the advertisement.
Below is a detailed breakdown of the planned activities.

\subsubsection{Introduction (5 min)}
The organizers will frame the workshop's theme and introduce the "remote collaborator" vision for the new paradigms of partnership between humans and LLM agents.

\subsubsection{Keynote (30 min)}
We will feature a keynote by a leading scholar whose work bridges HCI, CSCW, and AI in the context of human-agent collaboration. 
The talk will provide a theoretical anchor for our discussions, reflecting on foundational principles of collaboration and how they might be challenged or extended in an era of AI agents as equal partners. 
The talk is expected to span 25 minutes with 5 minutes for Q\&A. 

We have reached out to three keynote speakers, who are supportive of joining our workshop, but pending the acceptance of this proposal and their availability. 
They are Diyi Yang from Stanford University, who would be able to share insights on the intersection of technical development and computational social science;
Michael Bernstein from Stanford University, who would offer a perspective focusing on LLM-based social simulacra and the role of HCI in social science research; 
and Marshall Van Alstyne from Boston University, who could provide a unique perspective from economics and information systems on the practical implications of human-agent collaboration.

\subsubsection{Panel Discussion (20 min)} 
Following the keynote, we will host a 20-minute interactive panel to facilitate direct conversation between multi-disciplinary experts and attendees.
The panel consists of four panelists and one moderator. 
We will invite scholars from diverse research backgrounds, such as social science, AI, and design. The moderator will prompt the discussion with a single guiding question, for example: \textit{"What is the significant barrier to designing LLM agents that collaborate with humans in a mutual partnership?"} The floor will then immediately open to the audience to ask questions of specific panelists. This interactive format allows participants to identify shared challenges and contrast diverse perspectives early in the workshop agenda.

\subsubsection{Featured Lightning Talks (35 min)}
After the panel discussion, we will feature a series of lightning talks. This session combines three to four \textbf{invited talks from leading researchers} with three to four \textbf{authors of selected workshop submissions}. Each presentation will last 4 minutes, followed by a joint 5-minute Q\&A session at the end of the block.

Invited talks will offer a comprehensive overview of recent research progress on topics including LLM agent design, human-agent collaboration, and downstream applications. 
These talks are conceptual provocations designed to seed the workshop with conflicting ideas, setting an energetic and intellectually stimulating tone. 
Confirmed presenters include Bingsheng Yao (Northeastern University) and Chaoran Chen (University of Notre Dame).

Integrating selected submissions into this session allows the broader community to showcase a variety of novel designs and empirical findings alongside established experts.
The program committee will select the spotlights with a focus on diverse applications and innovative methodologies that extend the boundaries of current research.
All remaining accepted submissions will be presented in the poster session.


\subsubsection{Poster Session (30 min)}
During the coffee break, we will have a poster session for participants to present their latest work and discuss their insights on human-agent collaboration with other participants.
We will split the posters into two groups so that half of the participants can walk around while the other half is presenting.  

\subsubsection{Exploratory Design Session: Storyboarding Human-Agent Interactions (60 min) }
At the beginning of this activity, organizers will introduce the design sessions and form sub-groups.
Participants will be divided into groups of 5-7 people each, based on their identified research expertise, to ensure diversity of expertise within each group.
A moderator from the organizing committee will be assigned to join each group to facilitate discussion.

Two sessions of 30 minutes each will be conducted, with each group receiving a scenario probe centered on a classic collaborative challenge but re-imagined for a human-agent team in each session. 
The goal is to storyboard a critical interaction moment, focusing on the design of the interaction itself. 
Probes will encourage thinking about both foundations and frontiers.
Probe examples include:
\begin{enumerate}
    \item \textbf{Trust \& Accountability:} ``An agent working asynchronously makes a critical mistake. Storyboard the interface that allows the human to investigate the provenance of the error and provide corrective feedback that avoids similar errors and improves the agent's future behavior.''
    \item \textbf{Workspace Awareness in Multi-Human, Multi-Agent (MHMA) Teams:} ``In a team with three humans and three agents, the agents are collaborating at high speed. Storyboard a novel awareness visualization that gives humans an intuitive, ambient sense of the agents' progress and confidence without causing information overload.''
    \item \textbf{Creative Partnership:} ``A human designer is stuck. The agent partner offers a completely unexpected design direction. Storyboard how the agent presents this idea and the creative negotiation between the collaborators.''
\end{enumerate}

\subsubsection{Group Presentations, Synthesis, and Closing (30 min)}
A representative from each group will deliver a brief presentation after the design sessions to showcase their most insightful storyboard concept in a 3-minute pitch. 
This collective presentation will serve as the foundation for a broader discussion on establishing best practices in human-agent collaboration design and evaluation.
The organizers will facilitate a concluding synthesis and outline the next steps for the post-workshop white paper and community building.

\subsection{Post-Workshop Plans}
\label{sec:post-workshop}
We will produce a white paper and repository of workshop materials because, even if we have the maximum number of participants, this is a much broader conversation that is sorely needed. Our publishing strategy will ensure that the best practices and other results of the workshop are widely disseminated and accessible for ongoing engagement and collaboration. 
Our plan consists of the following components:
\begin{enumerate}
    \item \textbf{An Agenda-Setting White Paper}: A white paper summarizing the workshop discussions will be drafted by the workshop organizers and participants post-workshop and submitted to a leading journal or conference, such as Communications of the ACM, Interactions, NeurIPS, or CHI.
    This document will synthesize the theoretical discussions, the design patterns from the sessions, and the key questions identified during the workshop. 
    The drafting process will be collaborative; an outline will be shared with all participants, who will be invited to co-author sections based on their expertise. 
    We will hold virtual follow-up meetings to ensure the paper represents the collective intelligence of the workshop and submit it to a high-impact venue.
    \item \textbf{Building a Sustained Community} 
    We see this workshop as the beginning of a sustained conversation.
    We will maintain the workshop website as a lasting online repository that includes all the resources used during the workshop, such as accepted papers, keynote slides (with the consent of the keynote speakers), research presentations, and the output of interactive sessions. 
    This repository will be accessible to the public and referenced in the white paper to encourage further exploration and experimentation with human-agent collaboration. 
    We will explore the possibility of forming a more formal Special Interest Group (SIG) and recurrently host the workshop at future HCI (e.g., CHI, UIST), NLP (e.g., ACL and NAACL), and ML (e.g., Neurips, AAAI) conferences, and use the same website to document these materials.
    \item \textbf{Plans to Publish Proceedings} 
    All accepted position papers will be compiled into a digital proceedings volume.
    This volume will be made publicly and permanently available on the workshop's website. 
    We will explore publishing the workshop proceedings in the ACM Digital Library. 
    To ensure archival stability and broad, citable access for the authors, we will also publish the proceedings on an established open-access platform such as arXiv.org or CEUR-WS.org. 
    This guarantees that the intellectual contributions of our participants are formally recognized and can be built upon by the broader research community.
\end{enumerate}

\section{Accessibility}

We are committed to providing an inclusive and accessible experience for all participants. 
To support this, we will work with the conference organizers to \textbf{ensure the physical workshop space is fully accessible} to individuals with mobility needs. 
We will guide all presenters to follow best practices for \textbf{content accessibility}, including using high-contrast color schemes in their slides and verbally describing key visual information. 
Furthermore, to allow participants to prepare and use their own assistive technologies, we will make \textbf{all digital materials}, such as accepted papers and presentation slides, \textbf{available on our workshop website}. 
Finally, we are prepared to \textbf{accommodate specific accessibility requests}. 
We will offer to arrange for services such as live captioning or sign language interpretation, and clear contact information for making these requests will be prominently displayed on our website.

\section{Call for Participation -- A Vision, Design Philosophy, and Empirical Framework for Achieving Successful Partnerships Between Humans and LLM Agents}
\label{sec:cfp}

Recent advances in Large Language Models (LLMs) open up possibilities for moving intelligent systems beyond the “tool” paradigm toward more equal partnerships with humans. These agents can engage in fluid natural language dialogue, exhibit role-playing behaviors, and demonstrate reasoning and planning capabilities. At the same time, they introduce risks such as misaligned autonomy, hallucinations, privacy and security vulnerabilities, and over-anthropomorphization.

This workshop calls on the HCI, CSCW, AI, and related communities to critically and creatively explore new paradigms of human–agent partnership. We invite contributions that draw on the rich history of technology-mediated human–human collaboration while also envisioning novel forms of mixed-initiative, multi-party, and asynchronous collaboration with AI agents.
Topics of interest include, but are not limited to:

\begin{itemize}
    \item Theoretical perspectives: applying and extending CSCW/HCI theories (e.g., common ground, workspace awareness, articulation work) to human–agent interaction
    \item Empirical studies of human–agent collaboration (e.g., trust calibration, coordination, failure recovery)
    \item Novel design paradigms for human–agent collaboration, including mixed-initiative workflows, coordination strategies, and asynchronous delegation
    \item Interface and interaction designs that make agents’ goals, actions, and limitations transparent to human partners
    \item Risks, challenges, and safeguards for collaboration (e.g., security, privacy, safety, misaligned autonomy)
    \item Critical perspectives on anthropomorphization, social accountability, and the ethics of agentic systems
\end{itemize}

\textbf{Submission Format}: We welcome 2–4 page submissions (ACM single-column format), including research articles, survey papers, and position papers that present original ideas, empirical studies, or design visions.

\textbf{Review Process:} Submissions will undergo a lightweight, non-anonymous review, focusing on the quality of ideas and diversity of perspectives. Accepted papers will be published online on the workshop website.

\textbf{Participation:} The workshop will be a 180-minute, in-person session featuring a keynote, panel discussion, invited lightning and spotlight talks, and collaborative design activities via storyboarding. 
All accepted papers will be posted on the workshop website and feature a poster presentation.
A number of selected submissions will be presented as community spotlight talks.
More details of our workshop and paper submission can be found at \textit{\href{https://chi26workshop-human-agent-collaboration.hailab.io/}{https://chi26workshop-human-agent-collaboration.hailab.io/} }

\noindent \textbf{Important Date:} \\
CFP released date: December 18, 2025 \\
Submission due date: February 15, 2026 \\
Notification of acceptance date: February 28, 2026 \\
Workshop date: April 13, 2026


\bibliographystyle{ACM-Reference-Format}
\bibliography{sample-base}


\end{document}